\documentclass{PoS}

\usepackage{dsfont}

\title{Thin and dressed Polyakov loops from spectral 
sums of lattice differential operators}

\ShortTitle{Thin and dressed Polyakov loops}

\author{Christian Hagen\footnote{Based on presentations by Erek Bilgici and
        Christian Hagen.} \hspace{1mm} and Falk Bruckmann\\
        Institut f\"ur Theoretische Physik, Universit\"at
        Regensburg, D-93040 Regensburg, Germany \vskip5mm}

\author{Erek Bilgici$^*$ and Christof Gattringer\\
        Institut f\"ur Physik, FB Theoretische Physik, Universit\"at Graz,
        A-8010 Graz, Austria\\ \\
        E-mails:
        \email{erek.bilgici@uni-graz.at, falk.bruckmann@physik.uni-r.de,
        christof.gattringer@uni-graz.at, christian.hagen@physik.uni-r.de}}

\abstract{
We represent thin and dressed Polyakov loops as spectral sums of eigenvalues of 
differential operators on the lattice. For that purpose we 
calculate complete sets of eigenvalues of the staggered Dirac and the
covariant Laplace operator for several temporal boundary conditions. 
The spectra from different boundary
conditions can be combined such that they represent single (thin) Polyakov
loops, or a collection of loops (dressed Polyakov loops). We analyze the role of the 
eigenvalues in the spectral sums below and above the critical temperature.}

\FullConference{The XXV International Symposium on Lattice Field Theory\\
                 July 30 - August 4 2007\\
                 Regensburg, Germany}

\begin{document}

\section{Motivation}

The phenomenology of QCD is governed by two prominent features,
confinement and spontaneous 
breaking of chiral symmetry. As one increases the temperature $T$ above some
critical value $T_c$, the theory becomes deconfined and chiral symmetry is
restored. This suggests that there could be a relation between the two
phenomena. Establishing or ruling out such a relation would be a major insight
into key mechanisms of QCD. 

\begin{figure}[t]
\begin{center}
\resizebox{0.78\textwidth}{!}{\includegraphics[clip]{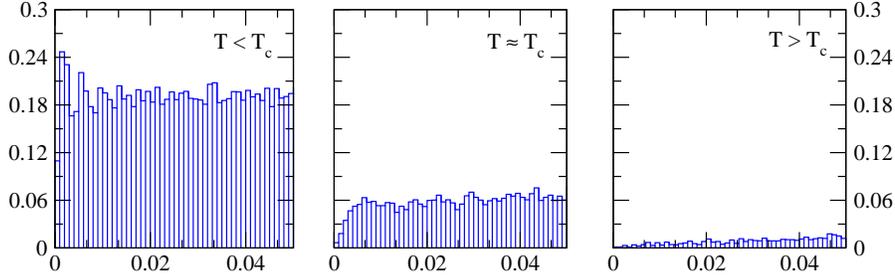}}
\end{center}
\caption{Histograms of the spectral density of the staggered lattice Dirac
  operator as a function of the modulus of the eigenvalues. Below the 
critical temperature the spectral density at the origin is non-zero. Near $T_c$ 
the density at the origin decreases and vanishes above the
critical temperature.}
\label{histo}
\end{figure}

The finite temperature transition of pure gauge theory, 
where the system changes from the confined $(T<T_c)$ 
into the deconfined phase $(T>T_c)$, 
can be understood as spontaneous breaking of the 
center symmetry \cite{McLerran:1981pb}, 
and the Polyakov loop is a suitable order parameter 
with $\langle P \rangle = 0$ for $T<T_c$ 
and $\langle P \rangle \neq 0$ for $T>T_c$ (see
Fig.~\ref{pol_cplane}). Concerning chiral symmetry breaking the order
parameter is given by the chiral condensate with 
$\langle \bar{\psi} \psi \rangle \neq 0$ for $T<T_c$ and 
$\langle \bar{\psi} \psi \rangle = 0$ 
for $T>T_c$. The chiral condensate in turn is related to the spectral density
of the Dirac operator by the Banks-Casher formula \cite{Banks:1979yr}:
$\langle \bar{\psi} \psi \rangle = - \pi \rho(0)$,
where $\rho(0)$ is the spectral density at the origin. 
When increasing the temperature above 
the critical value, the spectral density at the origin vanishes 
and with it the chiral condensate (see Fig. \ref{histo}). 

In a series of recent papers, \cite{Gattringer:2006ci}-\cite{Soeldner}, the
Polyakov loop, or other quantities that serve as order parameters for the
breaking of the center symmetry, were related to spectral sums of differential
operators on the lattice, in particular the Dirac and the covariant Laplace
operators. In this way also confinement is related to spectral quantities, as
is chiral symmetry breaking through the Banks Casher formula.
Since eigenvalues can be divided into IR and
UV regions in a natural way, the spectral representations allow one to analyze
whether confinement is dominated by IR or UV modes. In this paper we review
and extend our contributions to this enterprise.

\section{Derivation of spectral sums}

\subsection{Preliminaries}

\noindent
The thin Polyakov loop averaged over space is given by
\begin{equation}
P \; = \; \frac{1}{V_3} \sum_{\vec{x}} L(\vec{x}) \quad \mbox{ with } \quad 
L(\vec{x})=Tr_c\left[ \prod_{x_4=1}^{N_t} U_4(\vec{x},x_4) \right] \; ,
\label{gluonicloop}
\end{equation}
where $L(\vec{x})$ is a straight line of links in temporal direction 
closed around the periodic boundary and
$Tr_c$ denotes the color trace. 
Obviously, this quantity has a rather singular support. 
Due to this fact it is known to
have a poor continuum limit with large renormalization effects.

To address this problem, below we will define a new observable $P^{(q)}$
which we refer to as the 
\textit{dressed Polyakov loop with winding number $q$}. It is a collection of
loops all with the same number $q$ of windings around compactified time. 
So, for example, $P^{(1)}$ is a sum of the thin Polyakov
loop and all other loops winding once around the lattice in time direction,
where the latter loops may have arbitrary 
complicated detours in spatial directions and are weighted proportional to
their length. 

In the following two sections we derive the spectral sums for thin and dressed
Polyakov loops. For this purpose it is convenient to summarize our
conventions: The staggered Dirac operator is given by
\begin{eqnarray}
\label{stag}
D(x,y) &=& m \delta_{x,y} + \frac{1}{2} \sum_{\mu=1}^4 \eta_\mu(x) 
\left[ U_\mu(x) \delta_{x+\hat{\mu},y} - U_\mu(x - \hat{\mu})^\dagger 
\delta_{x-\hat{\mu},y} \right], 
\end{eqnarray}
where $\eta_\mu(x) = (-1)^{x_1+x_2+ \ldots + x_{\mu-1}}$ is the staggered
phase. The covariant Laplace operator on the lattice is defined as
\begin{eqnarray}
\label{laplace}
\Delta(x,y) &=& ( 8 + m^2 ) \delta_{x,y} - \sum_{\mu=1}^4 \left[ U_\mu(x) 
\delta_{x+\hat{\mu},y} + U_\mu(x - \hat{\mu})^\dagger \delta_{x-\hat{\mu},y} \right].
\end{eqnarray}
Below we will use the staggered Dirac operator for representing the thin
Polyakov loop, while the dressed loops are constructed from the Laplace
operator. We stress, however, that it is an easy exercise to generalize our
formulas, such that both the thin and the dressed Polyakov
loops can be obtained from both differential operators. The derivation can be
extended further to arbitrary lattice operators with
only nearest neighbor interaction (such as Domain Wall Fermions). 
The advantage of these
operators is that they are numerically cheap.

\begin{figure}[t]
\begin{center}
\resizebox{0.78\textwidth}{!}{\includegraphics[clip]{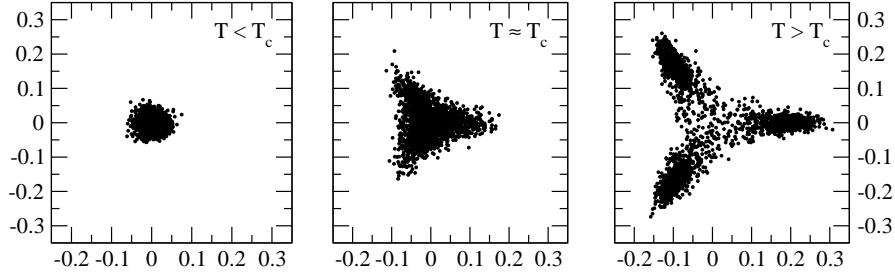}}
\end{center}
\caption{Scatter plot for the thin Polyakov loop in the complex plane for an
  ensemble of quenched SU(3) configurations.
As the temperature increases (left to right) the Polyakov loop acquires a
non-vanishing expectation value.}
\label{pol_cplane}
\end{figure}
\vspace{10mm}

\subsection{Thin Polyakov loops}
\label{thinderiv}

\noindent
In order to connect the thin Polyakov loop to spectral
sums we follow the derivation in 
\cite{Gattringer:2006ci}. We consider powers of the staggered Dirac operator 
(\ref{stag}). The 
$n$-th power of the operator then contains paths up to length $n$, dressed
with a product of $n$ links. For the special case where we consider the
operator to the power $N_t$, where $N_t$ is 
the time extent of our lattice, we obtain
\begin{eqnarray}
Tr_c D^{N_t}(x,x) & \; = \; & 
\frac{1}{2^{N_t}}Tr_c \prod_{s=1}^{N_t} U_4(\vec{x},s) \, - \,
\frac{1}{2^{N_t}}Tr_c \prod_{s=0}^{N_t-1} U_4(\vec{x},N_t-s)^\dagger
\\ \nonumber
& & + \, \mbox{ other loops, trivially closed} \\ 
\label{dnt}&=& \frac{1}{2^{N_t}} \big[ \, L(\vec{x}) \, - \, 
L^*(\vec{x}) \, \big] \, + \,  
\mbox{ other loops, trivially closed} .
\end{eqnarray}
From the set of all loops that contribute we have singled out the thin
Polyakov loop and its complex conjugate. These two straight loops are the only
ones that can close around compactified time. All others are 'trivially
closed', i.e., they do not wrap around the lattice in temporal direction.

The key observation is that a change of the temporal boundary conditions
\begin{equation}
\label{bcchange}
U_4(\vec{x},N_t) \; \rightarrow \; z\,U_4(\vec{x},N_t) \; , \quad |z|=1 \; ,
\end{equation}
where $z$ is a complex phase, only affects the Polyakov loops
\begin{equation}
L \; \longrightarrow \; z\, L \;,
\end{equation}
while the trivially closed loops remain unchanged. 
After such a change the expression (\ref{dnt})
reads
\begin{equation}
Tr_c D^{N_t}_z(x,x) \; = \; \frac{1}{2^{N_t}} 
\big[ \,  zL(\vec{x}) \, - \, z^*L^*(\vec{x}) \, \big] \, + \mbox{ other loops},
\end{equation}
where $D_z$ is the Dirac operator on a configuration where the 
temporal boundary conditions
are changed by a factor $z$. Averaging this expression over space and time we obtain
\begin{equation}
Tr\; D^{N_t}_z \; = \; \frac{V_4}{2^{N_t}}( z P - z^*P^* + X ) \; ,
\end{equation}
where $X$ is the sum of all trivially closed paths. 
We make use of this behavior to cancel 
the unwanted trivial contributions and project onto the 
thin Polyakov loop by taking linear combinations (coefficients $a_i$)
of this expression for three different boundary 
conditions $z_1,z_2$ and $z_3$
\begin{eqnarray}
P &\; \stackrel{!}{=} \; & 2^{N_t} \sum_{i=1}^3 a_i Tr \, (D^{N_t}_{z_i}) 
\nonumber 
\\
&=& \; P \;\underbrace{(z_1a_1 + z_2a_2 + z_3a_3)}_{\stackrel{!}{=}1} \, - \,  
P^*\underbrace{(z_1^*a_1 + z_2^*a_2 + z_3^*a_3)}_{\stackrel{!}{=}0} \, + \,
X\underbrace{(a_1+a_2+a_3)}_{\stackrel{!}{=}0} \;.
\end{eqnarray}
This leads to a set of linear equations for the coefficients $a_1,a_2$ and $a_3$
\begin{equation}
\left(
\begin{array}{ccc}
z_1 & z_2 & z_3 \\
z^*_1 & z^*_2 & z^*_3 \\
1 & 1 & 1
\end{array}
\right)
\left(
\begin{array}{c}
a_1 \\
a_2 \\
a_3
\end{array}
\right) \; = \;  
\left(
\begin{array}{c}
1 \\
0 \\
0
\end{array}
\right).
\end{equation}
One can show that this equation has a unique 
solution as long as the boundary conditions 
$z_1,z_2$ and $z_3$ are different.
For the particular choice $z_1=1$ and $z_2 = z_3^* = 
e^{\,i\frac{2\pi}{3}} \equiv z$ the thin 
Polyakov loop is then given by
\begin{eqnarray}
P &=& \frac{2^{N_t}}{V_4} \left[ \sum_i \left( \lambda^{(i)} \right)^{N_t} + \,
z^* \sum_i \left( \lambda_z^{(i)} \right)^{N_t} + \,
z \sum_i \left( \lambda_{z^*}^{(i)} \right)^{N_t} \right] \; ,
\label{finalthin}
\end{eqnarray}
where the $\lambda_z^{(i)}$ are the eigenvalues of 
the staggered Dirac operator for a given 
boundary condition $z$. Similar expressions can be obtained for other 
differential operators on the lattice.

\subsection{Dressed Polyakov loops}
\label{dressedderiv}

\noindent
In order to obtain the spectral sum for the dressed Polyakov loops we now use
the Laplace operator and apply a different strategy than used for the thin
loops. First, we rewrite the covariant Laplace operator on the lattice from 
Eq.~(\ref{laplace}) in the 
following way
\begin{equation}
\Delta \; = \; \frac{1}{\kappa} \, [ \,
  \mathds{1} \, - \, \kappa \, H \, ] \; \; , \; \; \;
H(x,y) \; = \; \sum_{\mu = 1}^4 \Big[ U_\mu(x) \, \delta_{x+\hat{\mu},y} + 
U_\mu(x-\hat{\mu})^\dagger  \, \delta_{x-\hat{\mu},y}
\Big] \; ,
\label{hoppmat}
\end{equation}
where the hopping matrix $H$ collects all terms 
which connect nearest neighbors on the 
lattice and the hopping parameter $\kappa$ is related to the bare mass $m$ via 
$\kappa = (8 + m^2 )^{-1}$. The inverse Laplace operator 
$\Delta^{-1}(x,x)$ can be expressed as geometric series in terms of powers of $H$,
\begin{equation}
 \Delta^{-1}(x,x) \; = \; 
\sum_{j=0}^\infty \kappa^{j+1} \, H^j (x,x) \; .
\label{loopderiv1}
\end{equation}
Since the hopping matrix $H$ contains only terms that connect 
nearest neighbors, the powers 
$H^j (x,x)$ in (\ref{loopderiv1}) correspond to closed loops of length $j$,
again dressed with the corresponding products of the link variables. Thus we can 
organize $\Delta^{-1}$ in terms of loops and order these with 
respect to their winding around 
time. We obtain (after taking the trace)
\begin{equation}
\mbox{Tr} \, \Delta^{-1} \; = \; \sum_x
\mbox{Tr}_c \, \Delta^{-1}(x,x) \; = \; \sum_{n \in \mathds{Z}} \!
\sum_{\;\;l \in {\cal L}_n} 
\kappa^{|l|+1} \; \mbox{Tr}_c \!\! \prod_{(y,\mu) \in l} \!\! U_\mu(y) \; ,
\label{loopderiv2}
\end{equation}
where ${\cal L}_n$ is the set of loops that wind exactly $n$-times around the 
time direction and $|l|$ denotes the length of a loop $l$. It is important
that now the sum on the right hand side contains loops of arbitrary length,
with all possible numbers of windings. To disentangle the different 
winding numbers we 
introduce U(1)-valued temporal boundary conditions for the Laplace operator 
which are most conveniently implemented by the replacement
\begin{equation}
U_4(\vec{x},N_t) \; \longrightarrow \; e^{i \varphi} \, U_4(\vec{x},N_t) \; ,
\label{u1bc}
\end{equation}
which corresponds to a parameterization $z=e^{i \varphi}$ in
(\ref{bcchange}). Each loop acquires 
a phase factor $e^{i \varphi n}$, corresponding to its winding number $n$:
\begin{equation}
\mbox{Tr} \, \Big(\Delta_\varphi \Big)^{-1} 
\; = \; \sum_{n \in \mathds{Z}} e^{i\varphi n}
\sum_{\;\;l \in {\cal L}_n} 
\kappa^{|l|+1} \; \mbox{Tr}_c \!\! \prod_{(y,\mu) \in l} \!\! U_\mu(y) \; .
\nonumber
\end{equation}
Integrating over $\varphi$ with a factor of 
$e^{-i \varphi q}$ projects to winding number 
$q$ and we end up with the dressed Polyakov loop $P^{(q)}$
(which we normalize with $V = N^3 N_t$),
\begin{equation}
P^{(q)} \; \equiv \; \frac{1}{V} \, \int_{-\pi}^\pi 
\frac{d \varphi}{2\pi} \, e^{-i \varphi q} \,
\mbox{Tr} \, 
(\Delta_\varphi)^{-1} 
\; = \; \frac{1}{V} \!\! 
\sum_{\;\;l \in {\cal L}_q} 
\kappa^{|l|+1} \; \mbox{Tr}_c \!\!\!\! \prod_{(y,\mu) \in l} \!\! U_\mu(y)
\, . 
\label{projectedloops}
\end{equation}
After the projection the loops still can be arbitrarily long but all have the
same winding number $q$. The individual loops are weighted with a factor
$\kappa^{|l|+1}$, where $|l|$ is the length of the loop. This leads to
an exponential suppression of long loops and the rate of the suppression is
determined by the value of
$\kappa$. Nevertheless arbitrary long loops may contribute and 
the formula (\ref{projectedloops}) is a sensible definition of dressed loops.

The final step to make the connection to spectral properties of the staggered
Dirac operator is to write the trace over $\Delta^{-1}$ as a spectral sum giving
\begin{equation}
P^{(q)} \; = \; \frac{1}{V}  \int_{-\pi}^\pi
\frac{d \varphi}{2\pi} \, e^{-i \varphi q} \,
\sum_{j=1}^{3V} \frac{\kappa}{1 \, - \, \kappa \, \lambda_{j}^{(\varphi)}}
\; = \; \frac{1}{V} \!\! 
\sum_{\;\;l \in {\cal L}_q} 
\kappa^{|l|+1} \; \mbox{Tr}_c \prod_{(y,\mu) \in l} \!\! U_\mu(y)
\, ,
\label{finallaplace}
\end{equation}
where $\lambda_{j}^{(\varphi)}$ is the 
$j$-th eigenvalue of the hopping matrix $H$ (\ref{hoppmat}) for boundary 
condition angle $\varphi$. The dressed 
Polyakov loop transforms in the same way as the 
thin Polyakov loop. Thus the dressed 
Polyakov loops are proper order parameters for the
breaking of the center symmetry. They are also not as singular as the thin
Polyakov loops and thus are expected to have a better continuum limit with 
smaller renormalization effects. In an 
analogous way dressed Polyakov loops can be constructed 
also with the eigenvalues of the 
Dirac operator instead of the Laplace spectrum. 

We stress that both equations (\ref{finalthin}) and (\ref{finallaplace}) are
exact results that hold for individual gauge configurations. However, for a practical
evaluation with numerically generated gauge configurations the two formulas
have a different status. Equation (\ref{finalthin}) for the thin
loops holds to machine precision when the result of the spectral sum is
compared to the direct evaluation of the thin Polyakov loop from the gluonic
definition (\ref{gluonicloop}). The expression (\ref{finallaplace}) for the
dressed loop on the other hand contains an integral over a continuum of
boundary conditions. That this integral can indeed be approximated numerically 
in a sensible way will be demonstrated in the next section.  

\begin{table}[b]
\begin{center}
\begin{tabular}{|c|c|c|c|}
\hline
$\beta$      & $L^3\times N_t$                 & a[fm]    & T[MeV]  \\ \hline\hline
$7.00$	     & $6^3\times 4, 12^3\times 4$   & $0.351(3)$ & $140$   \\ \hline
$7.60$	     & $6^3\times 4, 12^3\times 4$   & $0.194(4)$ & $254$   \\ \hline
$7.91$	     & $6^3\times 4, 12^3\times 4$   & $0.146(2)$ & $336$   \\ \hline\hline
$7.40$	     & $             12^3\times 6$   & $0.234(2)$ & $140$   \\ \hline
$8.06$	     & $             12^3\times 6$   & $0.129(1)$ & $255$   \\ \hline
$8.40$	     & $             12^3\times 6$   & $0.098(1)$ & $337$   \\ \hline
\end{tabular}
\end{center}
\caption{Ensembles used in our simulations. For the $6^3\times 4$ 
lattices 2000 configurations are available, 
for $12^3\times 4$ and $12^3\times 6$ we have 
used 20 configurations.}
\label{tab1}
\end{table}

\section{Numerical analysis}

\subsection{Simulation details}

\noindent
For the numerical analysis of our spectral sums (\ref{finalthin}) 
and (\ref{finallaplace}) 
we use quenched $SU(3)$ gauge configurations generated with the
L\"uscher-Weisz action \cite{LuWeact}. We work on lattices with different 
sizes ranging from
$6^3\times4$ up to $12^3 \times 6$ at temperatures below and above the critical value
$T_c$ (see Table~\ref{tab1}). For those we compute complete spectra of
both the staggered Dirac 
operator and the hopping matrix $H$ of the Laplace operator for up to 
16 different boundary 
condition. The lattice spacing was set \cite{scale} with
the Sommer parameter $r_0=0.5fm$. All error bars we show are 
statistical errors determined
with single elimination jackknife.

\subsection{Thin Polyakov loops}
\label{thinresults}

\noindent
For the thin Polyakov loops we restrict 
ourselves to the three boundary conditions $z_1=1$ 
and $z_2 = z_3^* = e^{i\frac{2\pi}{3}} 
\equiv z$ which are used to derive the spectral
sum (\ref{finalthin}). 
A different choice would change our results only marginally.

\begin{figure}[t]
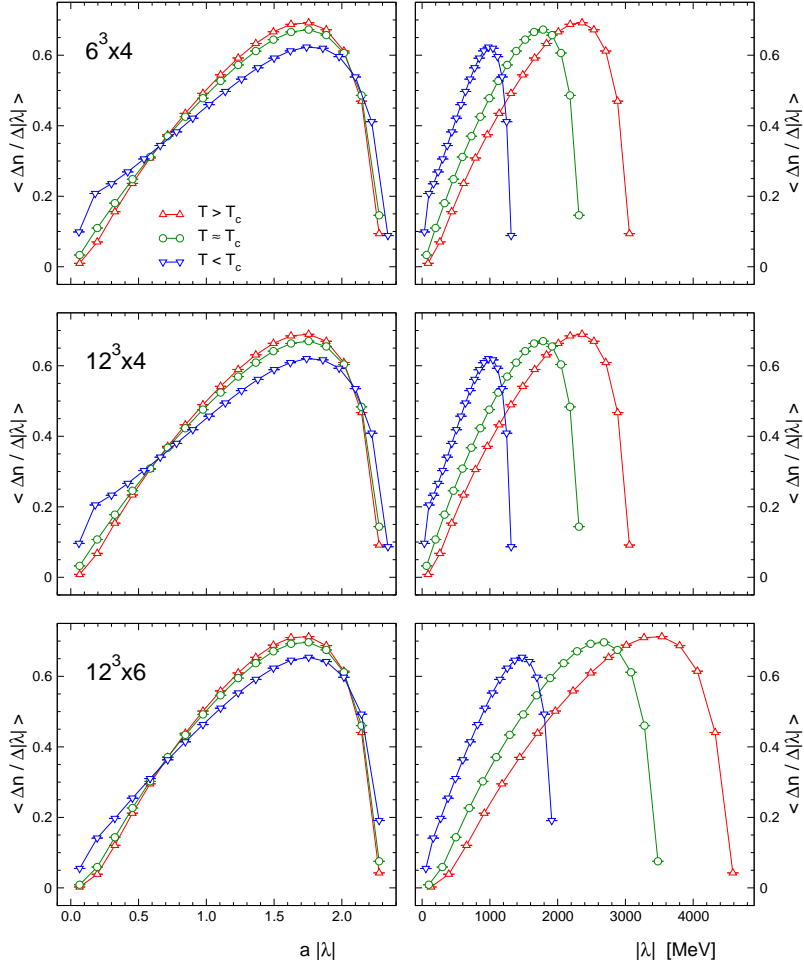

\begin{center}
\resizebox{0.7\textwidth}{!}{\includegraphics[clip]{evalcount_latt_6x4.eps} \hspace{4mm}
\includegraphics[clip]{evalcount_6x4.eps}}
\vskip3mm
\resizebox{0.7\textwidth}{!}{\includegraphics[clip]{evalcount_latt_12x4.eps} \hspace{4mm}
\includegraphics[clip]{evalcount_12x4.eps}}
\vskip3mm
\resizebox{0.7\textwidth}{!}{\includegraphics[clip]{evalcount_latt_12x6.eps} \hspace{4mm}
\includegraphics[clip]{evalcount_12x6.eps}}
\end{center}
\caption{Distribution of the eigenvalues of the staggered Dirac operator for different 
lattices and different temperatures. The l.h.s.~plots are in lattice units, 
while on the r.h.s.~we
use MeV.}
\label{evalcount}
\end{figure}

After computing all eigenvalues of the staggered Dirac operators we order them
w.r.t.~to their absolute value and organize them in bins. 
For each bin we calculate a number of 
observables. The first observable is the 
distribution of the eigenvalues, which we show in 
Fig.~\ref{evalcount} for different spatial 
and temporal extents of the lattice and for three 
different temperatures above, below and approximately at the critical value
$T_c$. The l.h.s.~set of those plots shows the distribution as a function of the
size of the eigenvalues in lattice units, while the r.h.s.~plots are in MeV. One can 
clearly see that the spectral density at the 
origin vanishes and chiral symmetry is restored as 
one increases the temperature above the critical value. The density of
eigenvalues reaches a maximum after about three quarters of the eigenvalues
and then quickly drops towards the UV cutoff. 

\begin{figure}[t]
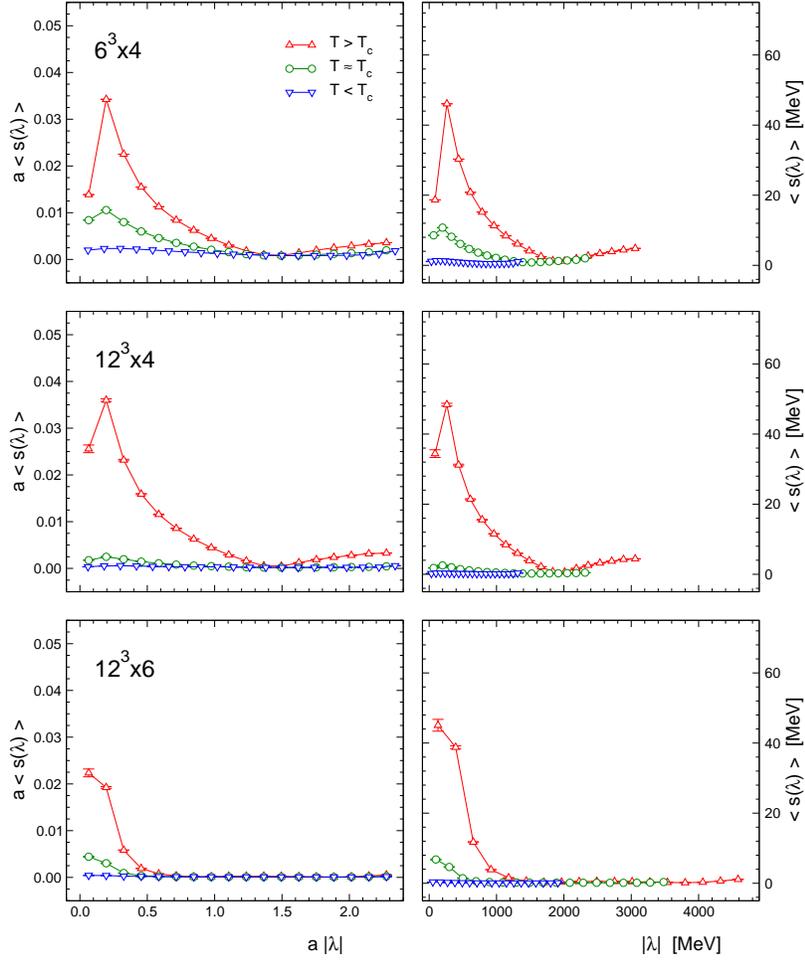

\begin{center}
\resizebox{0.7\textwidth}{!}{\includegraphics[clip]{shift_latt_6x4.eps} \hspace{4mm}
\includegraphics[clip]{shift_6x4.eps}}
\vskip3mm
\resizebox{0.7\textwidth}{!}{\includegraphics[clip]{shift_latt_12x4.eps} \hspace{4mm}
\includegraphics[clip]{shift_12x4.eps}}
\vskip3mm
\resizebox{0.7\textwidth}{!}{\includegraphics[clip]{shift_latt_12x6.eps} \hspace{4mm}
\includegraphics[clip]{shift_12x6.eps}}
\end{center}
\caption{Average shift of the eigenvalues calculated as defined in 
  Eq.~(3.1) as a function of $\lambda$. Again we use lattice units in the
  l.h.s.~set of plots and MeV on the r.h.s.}
\label{shift}
\end{figure}

Fig.~\ref{evalcount} is particularly important when one analyzes which part of
the spectrum contributes to the spectral sum (\ref{finalthin}) for the thin
Polyakov loop. For the interpretation of such an analysis it is necessary to
take into account the global distribution of the eigenvalues, and this is
exactly what Fig.~\ref{evalcount} provides.

In the spectral sum (\ref{finalthin}) the thin Polyakov loop emerges through a
shift of the eigenvalues as the boundary condition is changed. It is
interesting to analyze how strongly different eigenvalues of the
spectrum are shifted when the boundary condition is changed. We quantify this
question by studying the averaged shift of the eigenvalues given by 
\begin{equation}
\label{shiftformula}
s(\lambda) \; = \; \left(|\lambda-\lambda_z| \, + \, 
|\lambda-\lambda_{z^*}| \, + \, 
|\lambda_z-\lambda_{z^*}| \right)/3 \;.
\end{equation}
Since above the phase transition the thin Polyakov loop obtains a finite value while in 
the confined phase it is approximately zero, the shift of the eigenvalues should
change as one increases the temperature. This is exactly what we find in our
results for the average shift $s(\lambda)$ presented in Fig.~\ref{shift}). The
individual data points are obtained by averaging over all eigenvalues in a bin
and then over all configurations in the ensemble.

It is obvious, that over the whole range of the spectrum the average shift 
is almost zero below the critical temperature. For high temperatures, however, we find large
shifts for low-lying eigenvalues and little or no shift when the UV region is
approached. The small shift at the UV end of the spectrum disappears when
finer and larger lattices are used.

After understanding the global distribution of the eigenvalues and their shift
under a change of the boundary conditions, we can start to analyze the
individual contributions to the spectral sum (\ref{finalthin}). This
individual contribution is given by
\begin{equation}
c(\lambda) \; = \; \frac{2^{N_t}}{V_4} \left[ \left( \lambda \right)^{N_t} \, + \, 
z^* \left( \lambda_z \right)^{N_t} \, + \, z \left( \lambda_{z^*}
\right)^{N_t} \right] \; .
\end{equation}

\begin{figure}
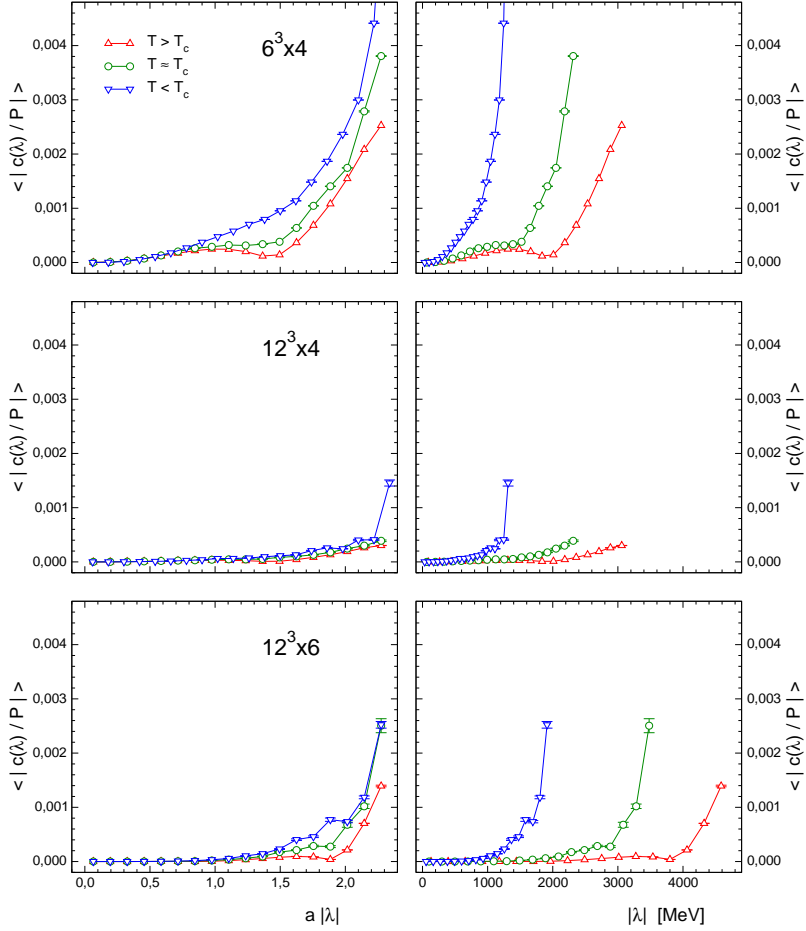

\begin{center}
\resizebox{0.7\textwidth}{!}{\includegraphics[clip]{contrib_latt_6x4.eps} \hspace{4mm}
\includegraphics[clip]{contrib_6x4.eps}}
\vskip3mm
\resizebox{0.7\textwidth}{!}{\includegraphics[clip]{contrib_latt_12x4.eps} \hspace{4mm}
\includegraphics[clip]{contrib_12x4.eps}}
\vskip3mm
\resizebox{0.7\textwidth}{!}{\includegraphics[clip]{contrib_latt_12x6.eps} \hspace{4mm}
\includegraphics[clip]{contrib_12x6.eps}}
\end{center}
\caption{Contribution (3.2) of individual eigenvalues to the spectral sum for the
thin Polyakov loop given by Eq.~(2.12). As before we use lattice units in the
l.h.s.~set of plots and MeV on the r.h.s. }
\label{contrib}
\end{figure}

Our results are shown in Fig. \ref{contrib}, where we plot the 
absolute value of the individual contributions after normalizing them with
$P$, i.e., we divide by the value of the thin
Polyakov loop. It is obvious that mainly the UV modes contribute to the thin
Polyakov loop. This observation is essentially independent of the temperature 
and the size of the lattice. This is a finding which is not a-priori
obvious, given the fact that it is the IR modes that show the largest shifts
in Fig.~\ref{shift}. On the other hand, the large power $N_t$ in the spectral
sum (\ref{finalthin}) drastically enhances the contributions of the 
UV eigenvalues.

Interesting are also the dips which form with increasing temperature at 
about $1.5a$ for $N_t=4$ and at $0.8a$ and $1.9a$ for $N_t=6$. It can be shown
that these dips are correlated with a change of the direction of the shift of
the eigenvalues for $z$-valued boundary conditions relative to the eigenvalues
computed with periodic boundary conditions.

\begin{figure}
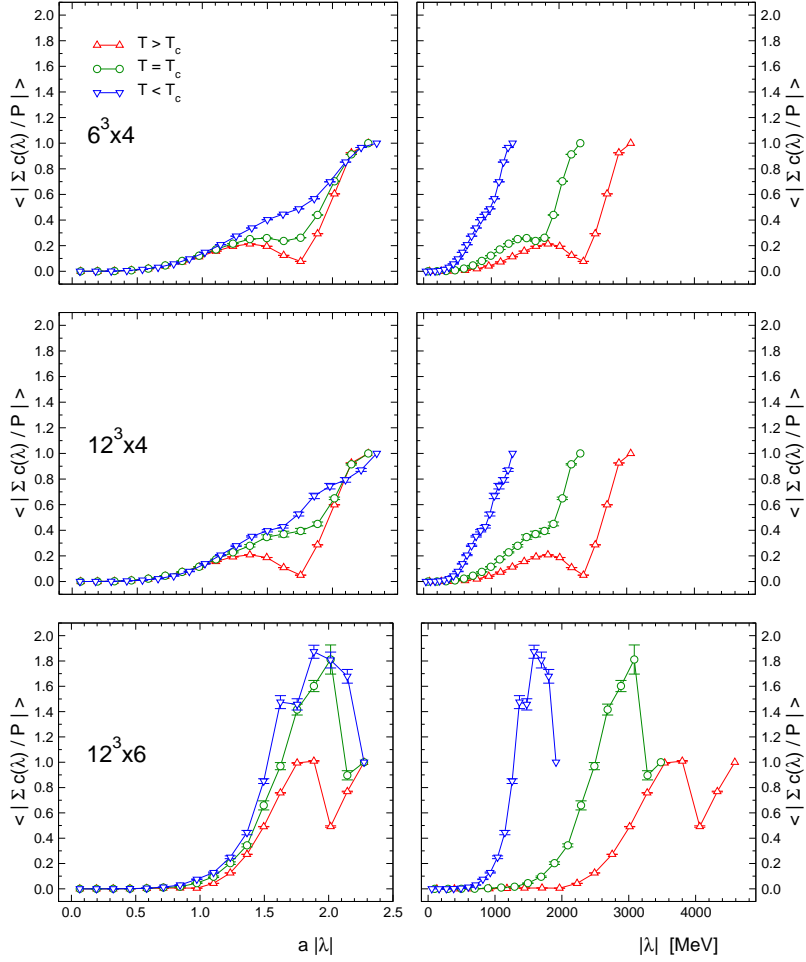

\begin{center}
\resizebox{0.7\textwidth}{!}{\includegraphics[clip]{accum_latt_6x4.eps} \hspace{4mm}
\includegraphics[clip]{accum_6x4.eps}}
\vskip3mm
\resizebox{0.7\textwidth}{!}{\includegraphics[clip]{accum_latt_12x4.eps} \hspace{4mm}
\includegraphics[clip]{accum_12x4.eps}}
\vskip3mm
\resizebox{0.7\textwidth}{!}{\includegraphics[clip]{accum_latt_12x6.eps} \hspace{4mm}
\includegraphics[clip]{accum_12x6.eps}}
\end{center}
\caption{Partial sums of the spectral formula (2.12), plotted as a function of
  the cutoff.}
\label{accum}
\end{figure}

The plot in Fig.~\ref{contrib} shows that the individual contributions are
biggest at the UV end of the spectrum. On the other hand, Fig.~\ref{evalcount}
shows that for the largest eigenvalues the density shows a sharp drop. It
is an interesting question which effect, size of contribution versus density,
wins out. This question can be addressed by considering only partial sums of
(\ref{finalthin}). In Fig.~\ref{accum}, we show such partial sums as a function of
the cutoff, i.e., as a function of the largest eigenvalue included in
(\ref{finalthin}). Again we divide by the Polyakov loop $P$, such that when 
all eigenvalues are summed, i.e., at the largest value of $\lambda$ in
Fig.~\ref{accum}, the curve approaches 1.

For the $12^3\times 6$ lattice, we find that for low and medium temperatures
the partial sums overshoot by nearly a factor of two
before they reach the correct value at the largest values of $\lambda$. Thus
the behavior of the partial sums is non-monotonic. On the other hand, 
for small temperatures the Polyakov loop should vanish. Thus the 
overshooting is not problematic, since the accumulated 
contributions stay approximately zero 
over the whole range of the spectrum. We also find that 
the thin Polyakov loop is recovered
only after all UV modes are included in the spectral sum. Again we see the
appearance of dips at exactly the same values which we have already identified
in Fig.~\ref{contrib}. 

\begin{figure}
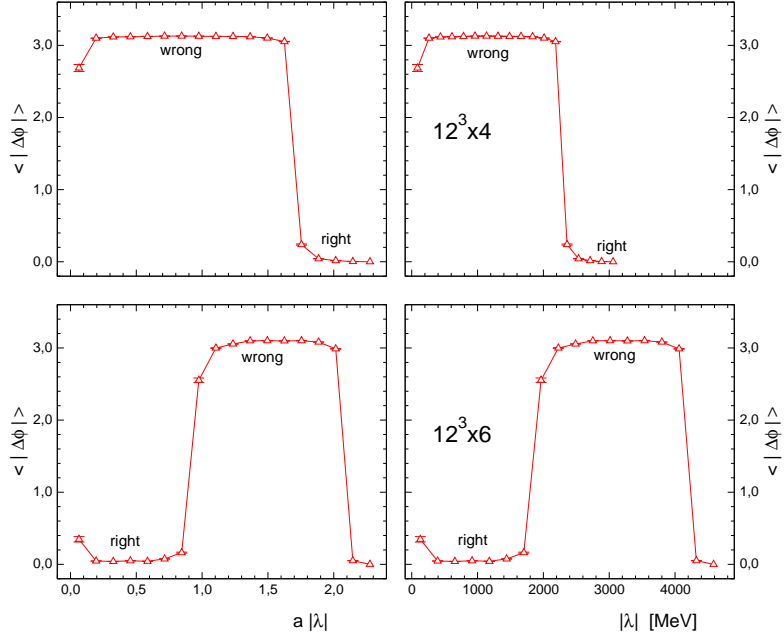

\begin{center}
\resizebox{0.68\textwidth}{!}{\includegraphics[clip]{phase_latt_12x4.eps} \hspace{4mm}
\includegraphics[clip]{phase_12x4.eps}}
\vskip3mm
\resizebox{0.68\textwidth}{!}{\includegraphics[clip]{phase_latt_12x6.eps} \hspace{4mm}
\includegraphics[clip]{phase_12x6.eps}}
\end{center}
\caption{
Phase shift of the truncated sums relative to the thin Polyakov loop for
$T>T_c$. The upper plots are for a $12^3\times 4$ lattice, while the lower plots are for 
$12^3\times 6$.}
\label{phase}
\end{figure}

So far we have only considered absolute values of quantities. However, the
Polyakov loop and also the spectral sums are complex numbers and we can
analyze their relative phase.
In Fig. \ref{phase} we plot the angle $\Delta \phi$ between the truncated sums  
and the resulting thin Polyakov loop as a function of the cutoff used in the
truncated sum. We show data for a $12^3\times 4$ and a $12^3\times 6$ lattice,
both at $T>T_c$. We find that on the former lattice 
the accumulated contributions start with the 
wrong sign and recover the correct sign at the end. 
On the other lattice, although starting and 
ending with the correct sign, there is an intermediate region in which the truncated 
sum has the wrong sign. A comparison with lattices of different spatial size but the
same temporal extent reveals, that this behavior depends only on the temporal extent.

A possible explanation for this phenomenon 
is the fact that the eigenvalues $\lambda$ of the staggered 
Dirac operator are purely imaginary. Thus $\lambda^{4n}$ is a real positive number, while
$\lambda^{4n+2}$ is real but negative. If we consider only the first term in the 
truncated sum we find:
\begin{eqnarray}
c(\lambda^{(1)}) &=& \frac{2^{N_t}}{V_4} \left[ \left( \lambda^{(1)}
  \right)^{N_t} + z^* \left( \lambda_z^{(1)} \right)^{N_t}+ z \left(
  \lambda_{z^*}^{(1)} \right)^{N_t} \right] 
\nonumber
\\
&=& \frac{2^{N_t}}{V_4} \left[ \alpha + z^* \tilde{\alpha}_z + z
  \tilde{\alpha}_{z^*} \right] \quad 
\stackrel{\tilde{\alpha}_{z^*} \approx \tilde{\alpha}_{z} \equiv
  \tilde{\alpha}}{=} \quad \frac{2^{N_t}}{V_4} \left[ \alpha + (z^* + z)
  \tilde{\alpha}\right] 
\nonumber
\\ 
&=& \frac{2^{N_t}}{V_4} \left[ \alpha - \tilde{\alpha}\right] 
\left\{
\begin{array}{lll}
< 0 & N_t = 4n   & \quad\mbox{as } \alpha,\tilde{\alpha} > 0 \\
> 0 & N_t = 4n+2 & \quad\mbox{as } \alpha,\tilde{\alpha} < 0
\end{array}
\right. ,
\end{eqnarray}
where we assume that the boundary conditions with 
phases $z$ and $z^*$ lead to approximately the same 
lowest eigenvalue $\tilde{\alpha}$, which is larger in size than the 
eigenvalue at periodic boundary conditions \cite{Kovacs,Synatschke:2007bz}.
Interestingly, this last relation between the relative sizes of the eigenvalues 
for non-trivial and periodic boundary conditions seems to change in certain 
regions in the spectrum of the staggered Dirac 
operator, thus creating the observed sign change in the truncated sums. The points where 
these sign changes occur depend only on the time extend of the lattice.

\subsection{Dressed Polyakov loops}
\label{dressedresults}

For the thin Polyakov loops the numerical analysis of the last section
demonstrates that the spectral sums for the thin loops are predominantly built up
from the UV modes. However, as we have already speculated above, part of the UV
dominance might come from the singular nature of the support of the thin
loop. Thus it is interesting to perform the same spectral analysis for the
dressed Polyakov loop. Such an analysis is planned for a future
publication. Here we focus on the feasibility of computing the integral over
the boundary angle in (\ref{finallaplace})
numerically and establish that the dressed Polyakov loop is
indeed an order parameter for the breaking of the center symmetry.

For the dressed Polyakov loops we compute complete spectra of $H$ for 16 values of the 
boundary angle $\varphi \in [0,\pi)$. The integrand in the spectral sum
(\ref{finallaplace}) is given by (omitting the normalization and the phase
$\exp(-i\varphi q)$)
\begin{equation}
S(\varphi) \; = \; \frac{1}{V}   \,
\sum_{j=1}^{3V} \frac{\kappa}{1 \, - \, \kappa \, \lambda_{j}^{(\varphi)}}
\; .
\label{integrand}
\end{equation}
Integrating $S(\varphi)$ over $\varphi$ with the phase $\exp(-i\varphi q)$
projects to the loops with winding number $q$. The important technical
question is how smooth a function of $\varphi$ this integrand is.  

\begin{figure}[t]
\hspace*{4mm}
\includegraphics[height=10.1cm,clip]{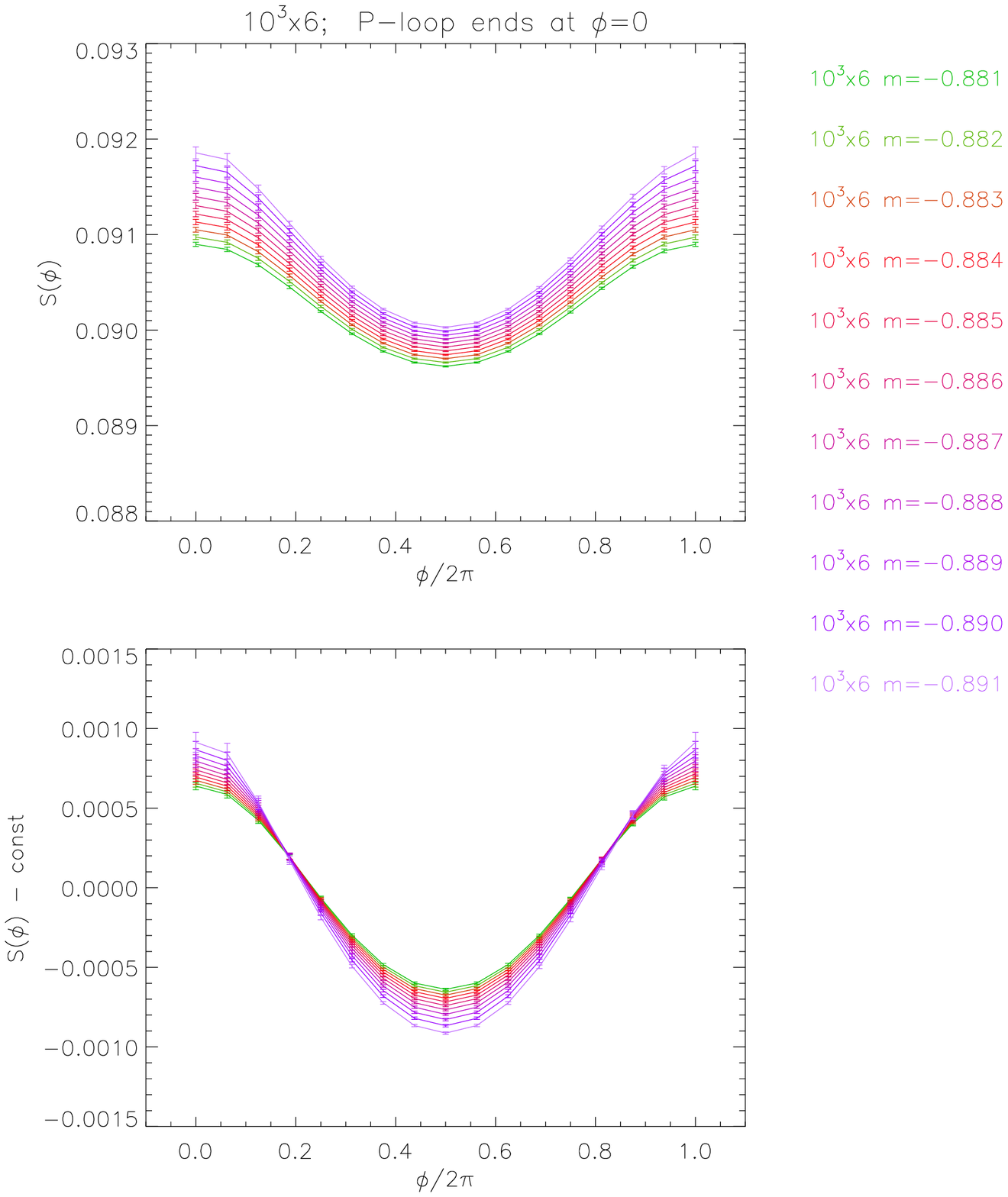}
\hspace{-4mm}
\includegraphics[height=10.1cm,clip]{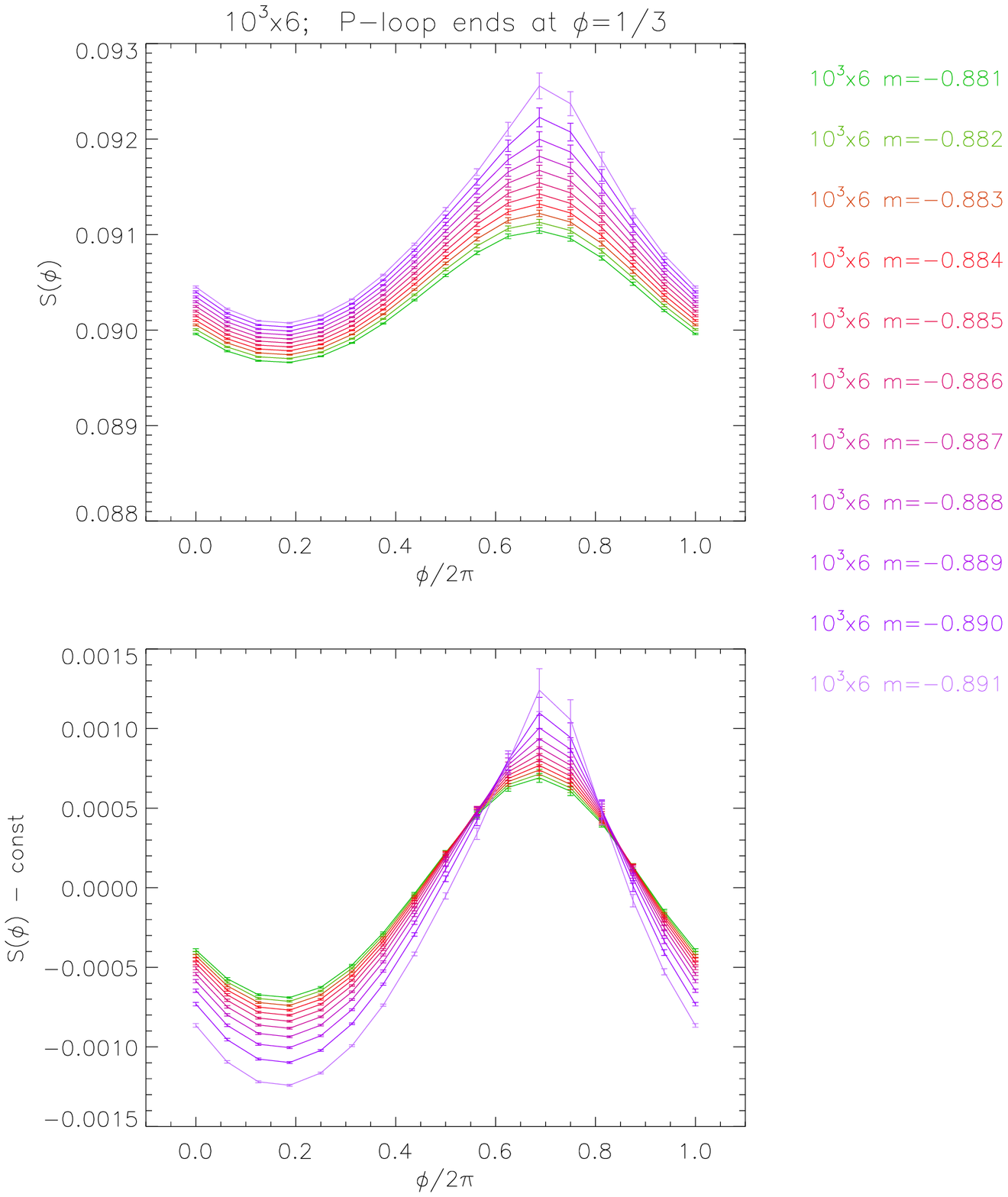}
\caption{
The spectral sum $S(\varphi)$ for the Laplace operator as a function of
$\varphi$. The l.h.s.\
plots are from an average over 20 configurations with real Polyakov loop, the
r.h.s.\ plots are for 20 configurations with a Polyakov loop with phase
$\sim \exp(i2\pi/3)$. In the top plots the full sum  $S(\varphi)$ is shown, while
in the bottom plots the constant $c = \int d \varphi S_L(\varphi)$ was subtracted.}
\label{figSDphi}
\end{figure}

Fig.~\ref{figSDphi} shows the integrand $S(\varphi)$ averaged over 20
configurations above $T_c$ with real Polyakov loops (l.h.s.~plots) and 20 configurations
with complex Polyakov loops (r.h.s.). The individual curves are for different
values of the mass parameter $m$ in the vicinity of the critical mass $m_c$
where the Laplace operator ceases to be invertible. In the bottom plots a
constant was subtracted. 

The plots show a clear cosine behavior for the ensemble with real Polyakov
loops (l.h.s.) and a shifted cosine for the complex ensemble (r.h.s.). For
low-lying modes a similar dependence on the phase at the boundary was observed
in \cite{lowphasedep}. 
It is obvious that dividing the interval $[0,\pi)$ into 16 sub-intervals should
be sufficient for a reliable numerical estimate of the $\varphi$-integral in the case 
of $q = 1$, where the phase factor is simply $cos \varphi -i \sin \varphi$. 
This is the most interesting case of dressed Polyakov loops with a single winding.  

\begin{figure}
\begin{center}
\resizebox{1.01\textwidth}{!}{
\hspace*{-7mm}
\includegraphics{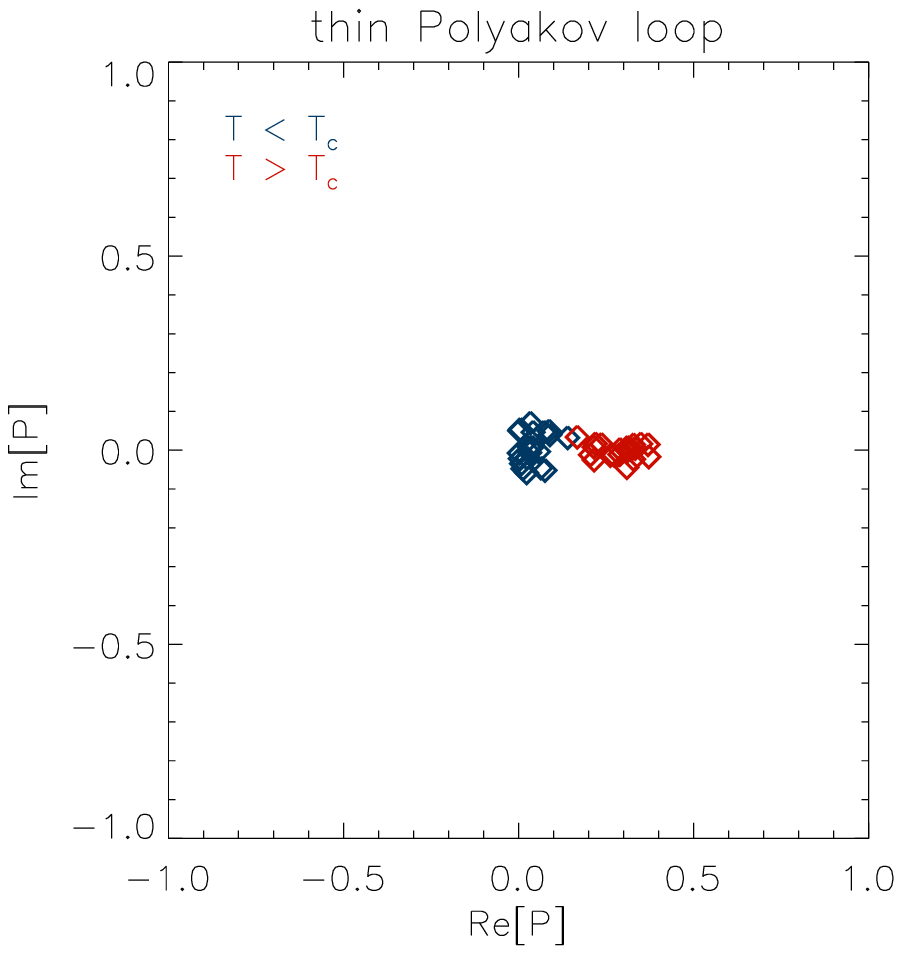}
\hspace{-3mm}
\includegraphics{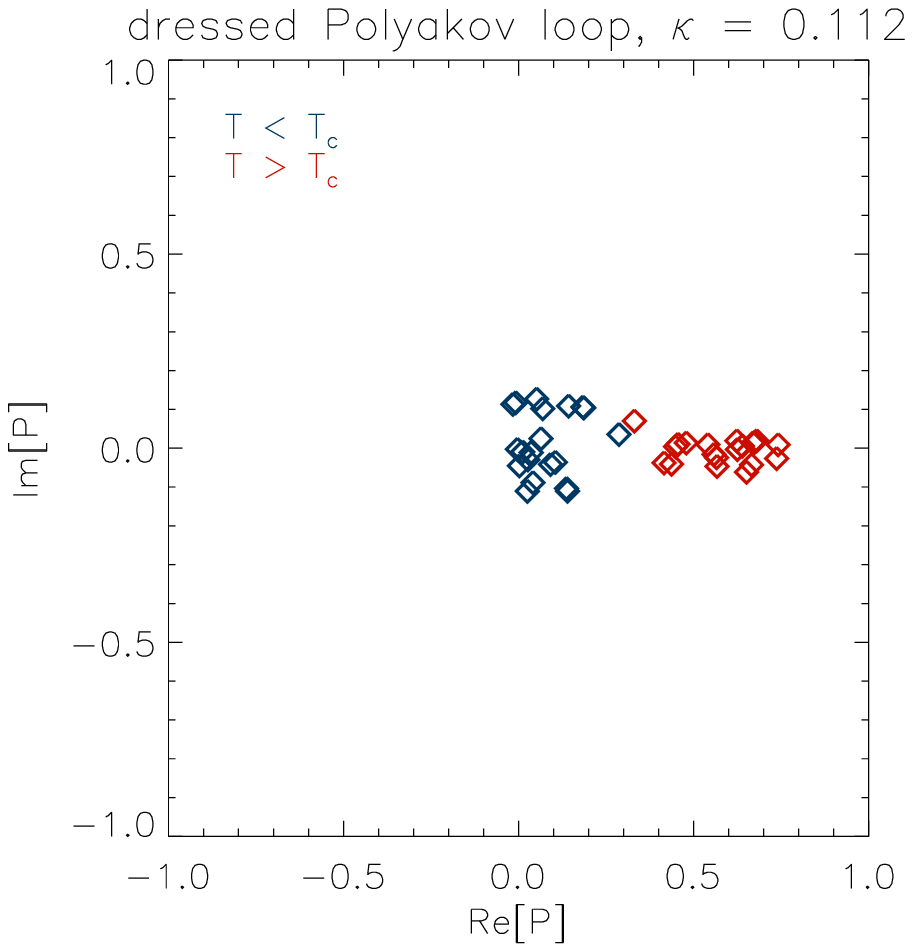}
\hspace{-3mm}
\includegraphics{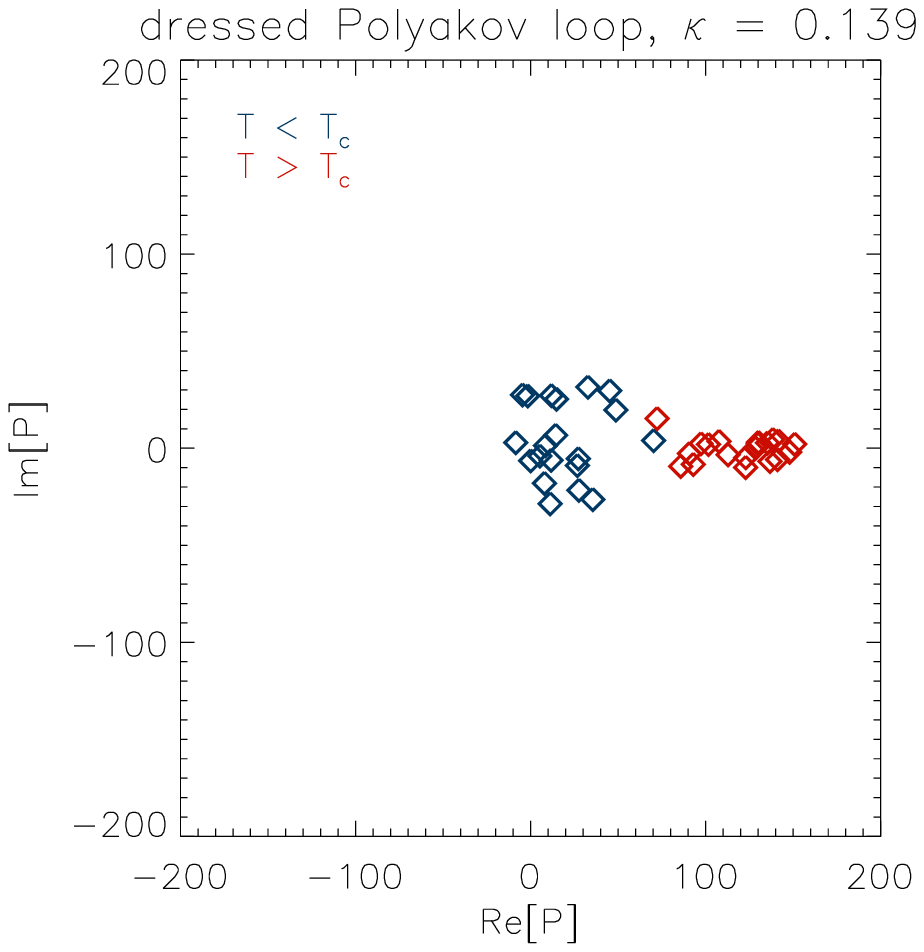}
}
\end{center}
\caption{Scatter plots of  thin (l.h.s.\ plot) 
and unscaled dressed Polyakov loops (center and l.h.s.\ plots) below (blue symbols)
and above $T_c$ (red symbols) for 20 configurations on a $12^3 \times 6$ lattice.} 
\label{scatter}
\end{figure}

Fig.~\ref{figSDphi} establishes that a numerical integration over $\varphi$ is
feasible. We implement this integration for the case of $q = 1$ using Simpson's rule.
The resulting dressed Polyakov loop $P^{(1)}$ transforms under center
transformations in the same way as the
thin Polyakov loop $P$ defined in (\ref{gluonicloop}). Thus it
should be an order parameter for breaking of center symmetry, but one even
expects that it shows a similar behavior as shown in Fig.~\ref{pol_cplane}
for the thin loop. In order to test this expectation we use 20 configurations 
below $T_c$ and 20 above $T_c$. The latter
all have a Polyakov loop which is essentially real, i.e., configurations
in the right leg of the distribution shown in the $r.h.s.$ plot of 
Fig.~\ref{pol_cplane}. On these two sets of configurations we compare the thin
loop with the dressed loop at two different values of $\kappa$. The
corresponding results in the complex plane are shown in Fig.~\ref{scatter}.

The plot shows clearly that for $T < T_c$ both, the thin and the dressed
Polyakov loops cluster around the origin, while above $T_c$ they develop a
non-vanishing value near the positive real axis. Even the pattern in the
distribution of the data points is similar for thin and dressed loops.

The analysis of Fig.~\ref{scatter} shows that the dressed Polyakov loop 
(\ref{finallaplace}) is a suitable order parameter for the breaking of center
symmetry. It is less singular and is expected to have a proper continuum
limit. We are currently analyzing the volume and temperature dependence of the
dressed loops and will then return to studying which part of the spectrum
predominantly contributes to the spectral sum for the dressed loops
(\ref{finallaplace}).

\acknowledgments

We thank Pierre van Baal, Tamas Kovacs, Kurt Langfeld, Wolfgang S\"oldner and
Andreas Wipf for stimulating discussions. Erek Bilgici has been supported by 
Fonds zur F\"orderung der wissenschaftlichen Forschung (FWF DK W1203-N08) and by 
NAWI-Graz. Christian Hagen has been supported by BMBF.

\end{document}